\documentclass[preprint,authoryear,3p,times]{elsarticle}
\usepackage{graphicx}
\usepackage{psfrag}
\usepackage{subfigure}
\usepackage{color}
\usepackage{url}
\usepackage{stfloats}
\usepackage{amsmath}
\usepackage{array}

\usepackage{amscd}
\usepackage{amssymb}

\begin{document}

\begin{frontmatter}

\title{Environment-dependent payoffs in finite populations}

\author{$Weihong~Xu^{a},~Yanling~Zhang^{a},~Guangming~Xie^{a}$}

\address{$^{a}$ Center for Systems and Control, State Key Laboratory for Turbulence and Complex Systems, College of Engineering, Peking University, Beijing 100871, China\\}

\begin{abstract}
In constant-payoff finite population games, when selection is weak
and population size is large, the one-third law serves as the
condition for a strategy to be advantageous. We generalize the
result to the case where payoff matrices are environment-dependent
and provide a more general law. In this way we model feedback from
the environment and show its impact on the dynamics.
\end{abstract}

\end{frontmatter}

\section{Introduction}
In the 1970s, J. Maynard Smith and G.R.Price brought the concept of
game into biology, and introduced the idea of evolutionary game
theory to describe population dynamics \citep{smith82,smith73}.
Since then, the idea has been widely used in biology, economy and
society, where how cooperation emerges among egoistic individuals is
a major concern.

Nowak et al. (2004) introduced the fixation probability of a single
A-individual (usually we consider a cooperator), denoted by
$\rho_A$, which is defined as the probability that its offspring
lineage invades and takes over a population of ($N-1$) many
B-individuals. If the fixation probability is bigger than that of a
neutral mutant, $1/N$, strategy A is deemed advantageous. Under weak
selection, computation has shown that $\rho_A>1/N$ for large $N$ if
\begin{equation}\label{one-third law}
a+2b>c+2d,
\end{equation}
where $a$, $b$, $c$ and $d$ are the payoffs of A versus A, A versus
B, B versus A and B versus B, or denoted with a payoff matrix
\begin{equation}
\begin{array}{@{}r@{}c@{}c@{}l@{}}
    & A & B  \\
    \left.\begin{array}
    {c} A \\B  \end{array}\right(
                    & \begin{array}{c} a \\ b \end{array}
                    & \begin{array}{c} c \\ d \end{array}
                          & \left)\begin{array}{c} \\  \\ \end{array}\right.
  \end{array}
  \end{equation}
This condition is equivalent of requiring the fitness of an A-player
to be higher than that of a B-player when the frequency of A is
$1/3$, and was dubbed the one-third law. Ohtsuki et al. (2007)
provided an intuition of the law.

In most game-theoretical studies, the payoff matrices are constant,
but some recent studies have been trying to model feedback from the
environment by varying the entries of the payoff matrices to the
state of the system \citep{tomochi02,lee11}.

In our paper, we treat finite population games with
environment-dependent payoffs as a generalization of the traditional
model. In the general case, we provide the condition for a strategy
to be advantageous, of which the one-third law is a specialization.
We then apply our findings to the prisoner's dilemma, the snowdrift
game (also called hawk-dove game or chicken game), and cases in
between, to show what effect feedback from the environment brings to
the evolution.

\section{Model and the condition for a strategy to be advantageous}

We discuss the situation where the payoff matrix
\begin{equation}
\begin{array}{@{}r@{}c@{}c@{}l@{}}
    & A & B  \\
    \left.\begin{array}
    {c} A \\B  \end{array}\right(
                    & \begin{array}{c} a(j) \\ b(j) \end{array}
                    & \begin{array}{c} c(j) \\ d(j) \end{array}
                          & \left)\begin{array}{c} \\  \\ \end{array}\right.
  \end{array}
  \end{equation}
varies with the number of A-players $j$, representing feedback from
the environment. For simplicity, $a(j)$, $b(j)$, $c(j)$ and $d(j)$
are assumed linear functions of $j$, i.e., $a(j)=a_0+a_1j$,
$b(j)=b_0+b_1j$, $c(j)=c_0+c_1j$, and $d(j)=d_0+d_1j$.

The model studied by Nowak et al. and Ohtsuki et al., where
$a_1=b_1=c_1=d_1=0$ \citep{nowak04,ohtsuki07}, can be considered a
special case. In Ohtsuki et al. (2007), it was shown that in the
limit of large $N$, one-third of the opponents that one meets (in
interactions leading to state changes) until either extinction or
fixation are A-players and two-thirds are B-players, and condition
\eqref{one-third law} is the exactly the same as the condition that
the average payoff of A-players along an invasion-path is higher
than that of B-players. In our generalized situation, though one
still meets B-players twice as often as A-players, the average
payoffs along an invasion-path cannot be calculated as easily.
Instead, we should take into consideration that $a$, $b$, $c$ and
$d$ vary with $j$. From
\begin{equation}\begin{aligned}
\frac{1}{N-1}\sum_{j=1}^{N-1}\bar{\sigma}_{1j} \left(
\begin{array}{cc}
(A\rightarrow A)|_ja(j) & (A\rightarrow B)|_jb(j) \\
(B\rightarrow A)|_jc(j) & (B\rightarrow B)|_jd(j) \\
\end{array}
\right)
\\=\frac{1}{N-1}\sum_{j=1}^{N-1}\frac{2(N-j)}{N} \left(
\begin{array}{cc}
(j-1)(a_0+a_1j) & (N-j)(b_0+b_1j)\\
j(c_0+c_1j) & (N-j-1)(d_0+d_1j) \\
\end{array}
\right),
\end{aligned}\end{equation}
where $\bar{\sigma}_{1j}$ is the relative frequency that state $j$
is visited \citep{ohtsuki07}, the average payoff of A-players along
an invasion-path is
\begin{equation}\begin{aligned}
\frac{1}{N-1}\sum_{j=1}^{N-1}\frac{2(N-j)}{N(N-1)}[(j-1)(a_0+a_1j)+(N-j)(b_0+b_1j)]
\\=\frac{1}{6(N-1)}[(\underbrace{a_1+b_1}_{A_2})N^2+(\underbrace{2a_0-a_1+4b_0+b_1}_{A_1})N+(\underbrace{-4a_0-2a_1-2b_0}_{A_0})],
\end{aligned}\end{equation}
and that of B-players is
\begin{equation}\begin{aligned}
\frac{1}{N-1}\sum_{j=1}^{N-1}\frac{2(N-j)}{N(N-1)}[j(c_0+c_1j)+(N-j-1)(d_0+d_1j)]
\\=\frac{1}{6(N-1)}[(\underbrace{c_1+d_1}_{B_2})N^2+(\underbrace{2c_0+c_1+4d_0-d_1}_{B_1})N+(\underbrace{2c_0-8d_0-2d_1}_{B_0})].
\end{aligned}\end{equation}
Therefore, for large $N$, the condition for strategy A to be
advantageous is
\begin{equation}\label{1}
A_2>B_2,
\end{equation}
or
\begin{equation}\label{2}
\left\{\begin{aligned}
A_2=B_2\\
A_1>B_1
\end{aligned}\right.,
\end{equation}
or
\begin{equation}\label{3}
\left\{\begin{aligned}
A_2=B_2\\
A_1=B_1\\
A_0>B_0
\end{aligned}\right..
\end{equation}
Here $a_1$, $b_1$, $c_1$ and $d_1$ are the main factors because $N$
is so large that on the path of invasion, for a significant amount
of time, $j$ is large enough that $a_1j\gg a_0$, similar for $b(j)$,
$c(j)$ and $d(j)$. When $a_1=b_1=c_1=d_1=0$, we obtain the one-third
law as a special case.

The above is verified with direct calculation of $\rho_A$.
\begin{equation}\label{5}
\rho_A=1/(1+\sum_{k=1}^{N-1}\prod_{j=1}^k\frac{g_i}{f_i})
\end{equation}
\citep{nowak04}, where $f_i$ and $g_i$ are fitness of strategies A
and B, respectively:
\begin{equation}
f_i=1-w+w[a(j)(j-1)+b(j)(N-j)]/(N-1)
\end{equation}
and
\begin{equation}
g_i=1-w+w[c(j)j+d(j)(N-j-1)]/(N-1),
\end{equation}
with $w$ ($w\ll1$) being the intensity of selection. Under weak
selection, calculations yield
\begin{equation}
\rho_A=\frac{1}{N}+\frac{\beta}{12N}[(A_2-B_2)N^2+(A_1-B_1)N+(A_0-B_0)]
\end{equation}
Thus for large $N$, the condition for $\rho_A>1/N$ is \eqref{1} or
\eqref{2} or\eqref{3}.

The payoff matrix of both the prisoner's dilemma and the snowdrift
game can be denoted with two parameters , the benefit $b$ and the
cost $c$ ($b>c$) \citep{fu09}. As an example, we study the case
where $b$ is a monotonically increasing linear function of the
number of cooperators $j$ in a group of fixed size $N$, i.e.,
$b(j)=p+qj$ ($p>1$ and $q>0$). For simplicity, we set $c=1$. This
arrangement can be understood that in a society where more
individuals cooperate, cooperation produces greater benefit, yet
there is also a stronger temptation to defect.

The prisoner's dilemma has the payoff matrix
\begin{equation}
\begin{array}{@{}r@{}c@{}c@{}l@{}}
    & C & D  \\
    \left.\begin{array}
    {c} C \\D  \end{array}\right(
                    & \begin{array}{c} b(j)-1 \\ b(j) \end{array}
                    & \begin{array}{c} -1 \\ 0 \end{array}
                          & \left)\begin{array}{c} \\  \\ \end{array}\right.
  \end{array}.
  \end{equation}
Thus, $A_2=a_1+b_1=q=c_1+d_1=B_2$, and we need to compare $A_1$ and
$B_1$. Here $A_1=2a_0-a_1+4b_0+b_1=2(p-1)-q+4(-1)+0=2p-q-6$, and
$B_1=2c_0+c_1+4d_0-d_1=2p+q+4\times0-0=2p+q$. Therefore $A_1<B_1$,
which suggests cooperation is disadvantageous (i.e., the fixation
probability of a single cooperator $\rho_C$ is smaller than $1/N$).

In contrast, the snowdrift game has the payoff matrix
\begin{equation}
\begin{array}{@{}r@{}c@{}c@{}l@{}}
    & C & D  \\
    \left.\begin{array}
    {c} C \\D  \end{array}\right(
                    & \begin{array}{c} b(j)-1/2 \\ b(j) \end{array}
                    & \begin{array}{c} b(j)-1 \\ 0 \end{array}
                          & \left)\begin{array}{c} \\  \\ \end{array}\right.
  \end{array}
  \end{equation}
Thus, $A_2=a_1+b_1=2q$, and $B_2=c_1+d_1=q$. Therefore $A_2>B_2$,
suggesting that cooperation is advantageous (i.e., $\rho_C>1/N$).

When $b$ is a constant, the one-third law yields $b-3>b$ for the
prisoner's dilemma, which is impossible, and $3b-3/2>b$ for the
snowdrift game, which is, however, satisfied for all $b$ (note that
we already require $b>1$). Therefore, cooperation is advantageous in
the snowdrift game but not in the prisoner's dilemma, which is still
the case, as we have shown, when $b$ is a monotonically increasing
linear function of $j$.

While this may not be very interesting so far, we further consider
the cases in between. There are situations not so bad as the
prisoner's dilemma and at the same time not as good as the snowdrift
game. We introduce a parameter $x$ ($0<x<1$), which is the closeness
of the payoff matrix to that of the prisoner's dilemma, to represent
those. And we use a linear combination of the payoff matrices of the
prisoner's dilemma and the snowdrift game to form our payoff matrix:
\begin{equation}
\begin{array}{@{}r@{}c@{}c@{}l@{}}
    & C & D  \\
    \left.\begin{array}
    {c} C \\D  \end{array}\right(
                    & \begin{array}{c} x[b(j)-1]+(1-x)[b(j)-1/2] \\ xb(j)+(1-x)b(j) \end{array}
                    & \begin{array}{c} -x+(1-x)[b(j)-1] \\ 0 \end{array}
                          & \left)\begin{array}{c} \\  \\
                          \end{array}\right.
  \end{array}.
  \end{equation}
$x$ and $1-x$ may also be understood as the probabilities that the
prisoner's dilemma and the snowdrift game are being played,
respectively, and our game is a combination of the two. $A_2>B_2$ (
i.e.,$a_1+b_1>c_1+d_1$) yields $qx+2q(1-x)>qx+q(1-x)$, i.e.,
$q(1-x)>0$, which is satisfied automatically. However, if $b$ is
constant, the one-third law \eqref{one-third law} requires
$(b-3)x+(3b-3/2)(1-x)>bx+b(1-x)$, i.e., $b\geq-3/4+3/{2(1-x)}$. This
is satisfied only when $b$ is big enough. In particular, when $x$
approximates $1$, this will require $b$ to approximate infinity.
Therefore, we arrive at an interesting conclusion: this increase of
$b$ with $j$ is helpful in establishing the dominance of cooperation
in cases between the prisoner's dilemma and the snowdrift game (or
cases that are combinations of the two). When $b$ is a monotonically
increasing function of $j$, a minimum portion of the snowdrift game
is needed to transform the prisoner's dilemma into a game where
cooperation is advantageous.

\section{Conclusion}
We derived the condition for strategy A to be advantageous in finite
population games, when selection is weak, population size is large,
and the payoff matrix varies with the number of A-players. When the
payoff matrix is constant, the condition becomes the one-third law
given by Nowak et al. (2004).

The condition was then applied to the study of the prisoner's
dilemma, the snowdrift game, and their combinations, whose payoff
matrices can all be denoted with two parameters, the benefit $b$ and
the cost $c$ \citep{fu09}. While an increase of $b$ with the number
of cooperators has little significance in reshaping the prisoner's
dilemma, where cooperation is disadvantageous, or the snowdrift
game, where cooperation is advantageous, it plays a rather important
role in cases that are combinations of the two.


\begin{thebibliography}{00}
\bibitem[Nowak et al., 2004]{nowak04} Nowak, M.A., Sasaki, A., Taylor, C., Fudenberg, D., 2004.
Emergence of cooperation and evolutionary stability in finite
populations. Nature 428, 646--650.

\bibitem[Smith et al., 1982]{smith82}
Smith, J.M., 1982. Evolution and the theory of games. J. Math. Biol.
54, 721--744.

\bibitem[Smith and Price, 1973]{smith73}
Smith, J.M., Price, G.R., 1973. The logic of animal conflict. Nature
246, 15--18.

\bibitem[Ohtsuki et al., 2007]{ohtsuki07}
Ohtsuki, H., Bordalo, P., Nowak, M.A., 2007. The one-third law of
evolutionary dynamics. J. Theor. Biol. 249(2), 289--295.

\bibitem[Tomochi and Kono, 2002]{tomochi02}
Tomochi, M., Kono, M., 2002. Spatial prisoner's dilemma games with
dynamic payoff matrices. Phys. Rev. E 65, 026112.

\bibitem[Lee et al., 2011]{lee11}
Lee, S., Holme, P., Wu, Z., 2011. Emergent hierarchial structures in
multiadaptive games. Phys. Rev. Lett. 106, 028702.

\bibitem[Fu et al., 2010]{fu09} Fu, F., Nowak, M.A.,
Hauert, C., 2010. Invasion and expansion of cooperators in lattice
populations: prisoner's dilemma vs. snowdrift games. J. Theor. Biol.
266 (3), 358--366.












\end{thebibliography}
\end{document}